\documentclass[11pt,twoside]{article}
\usepackage{asp2010}

\resetcounters

\begin{document}

\title{A dynamical study of the nova remnant of GK Per}
\author{T. Liimets,\altaffilmark{1,2}
  R.L.M. Corradi,\altaffilmark{3,4}
  M. Santander--Garc{\'{\i}}a,\altaffilmark{5,6}
  E. Villaver,\altaffilmark{7} 
  P. Rodr{\'{\i}}guez-Gil,\altaffilmark{3,4}
  K. Verro,\altaffilmark{1,2} and I. Kolka\altaffilmark{1}}
\altaffiltext{1}{Tartu Observatory, Observatooriumi 1, T\~oravere, 61602, Estonia}
\altaffiltext{2}{Institute of Physics, University of Tartu, Riia 142, 51014, Estonia}
\altaffiltext{3}{Instituto de Astrof{\'{\i}}sica de Canarias, V\'\i a L\'actea s/n, La Laguna, E-38205, Santa Cruz de Tenerife, Spain}
\altaffiltext{4}{Departamento de Astrof{\'{\i}}sica, Facultad de F\'\i sica y Matem\'aticas, Universidad de La Laguna, Avda. Astrof\'\i sico Francisco S\'anchez s/n, La Laguna, E-38206, Santa Cruz de Tenerife, Spain}
\altaffiltext{5}{Observatorio Astron\'omico Nacional, Ap.\ de Correos 112, E-28803, Alcal\'a de Henares, Madrid, Spain}
\altaffiltext{6}{Centro de Astrobiolog\'\i a, CSIC-INTA, Ctra de Torrej\'on a Ajalvir km 4, E-28850 Torrej\'on de Ardoz, Spain}
\altaffiltext{7}{Departamento de F\'{\i}sica Te\'orica, Universidad Aut\'onoma de Madrid, E-28049 Madrid, Spain}
\begin{abstract}
Due to their large expansion speed, the apparent growth of nearby
nova remnants such as GK Per can be easily resolved from ground-based
optical imagery on a timescale of months. If the expansion in the
plane of the sky is coupled with the Doppler shift velocities, an
almost complete dynamical picture is drawn. We will discuss our latest
results of such a study on GK Per.
\end{abstract}

\section{Introduction}
GK Per is remarkable in many aspects. It is the result of a bright
classical nova explosion in 1901. For the first time in astronomy,
superluminal motions were observed, and then interpreted as light
echoes.  The actual remnant of the explosion has turned out to be the
longest lived and most energetic classical nova remnant.

\section{Observations and Data Reduction}
Most of the imaging data was obtained with the Isaac Newton Telescope
(INT) between 2004 and 2011 using a narrow band $H\alpha$+[N II]
filter with a central wavelength of 6568~\AA\ and a bandpass of
95~\AA. The Wide Field Camera (WFC) with a pixel scale $0''.33$
pix$^{-1}$ was used.  Occasionally, the Nordic Optical Telescope (NOT)
with the Andalucia Faint Object Spectrograph and Camera (ALFOSC) with
a similar filter was used.  A total of 19 images, obtained with a
seeing between $0''.5$ to $1''.3$, were used for the data analysis.
To complement our data, five archival images were downloaded, observed
between 1987 and 1999 with a different set of telescopes and filters:
broad band R or narrow band $H\alpha$.

The long-slit spectra were obtained with the INT and NOT in 2007. 
At the INT the Intermediate Dispersion Spectrograph (IDS) was used with
grating R1200Y and a $1''.5$ slit covering a spectral range from
5730~\AA\ to 7610~\AA\ at a dispersion of 0.47~\AA~pix$^{-1}$.  At the
NOT ALFOSC with a grism~\#17 and $0''.5$ slit was used, giving
spectral range from 6350~\AA\ to 6850~\AA, and a dispersion of 0.26~
\AA~pix$^{-1}$. Observations were secured at twelve different position
angles, evenly distributed over the remnant. 

All data were reduced using IRAF (see more from \citet{tl_liim2012}).

\section{Imaging analysis}

\begin{figure}[!t]
 \begin{center}
 \includegraphics[width=6cm]{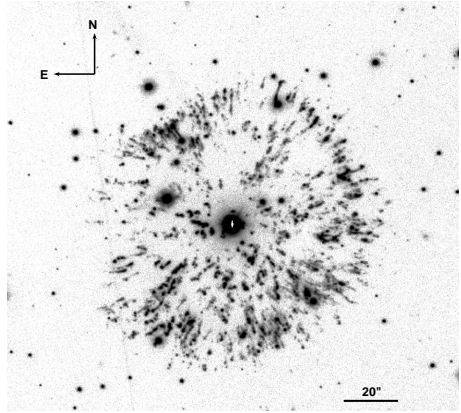}
\caption{NOT image of GK Per, obtained on September 2007. The field of view (FOV) is 2.8x2.8 arcmin$^2$.
\label{F-or}}
\end{center}
\end{figure}

The optical ejecta has a knotty roundish morphology (Fig.~\ref{F-or}).
However, some deviation from circular symmetry is evident. The
diameter of the remnant in the northeast-southwest (NE-SW) direction
is 10\% smaller than in northwest-southeast (NW-SE) direction.  The
remnant is composed of hundreds of knots and filaments (see also
\citet{tl_shara2012}) with different sizes and brightness. Many knots
have tails pointing away or towards the central source.

\begin{figure}[!t]
\begin{center}
\includegraphics[width=6.0cm]{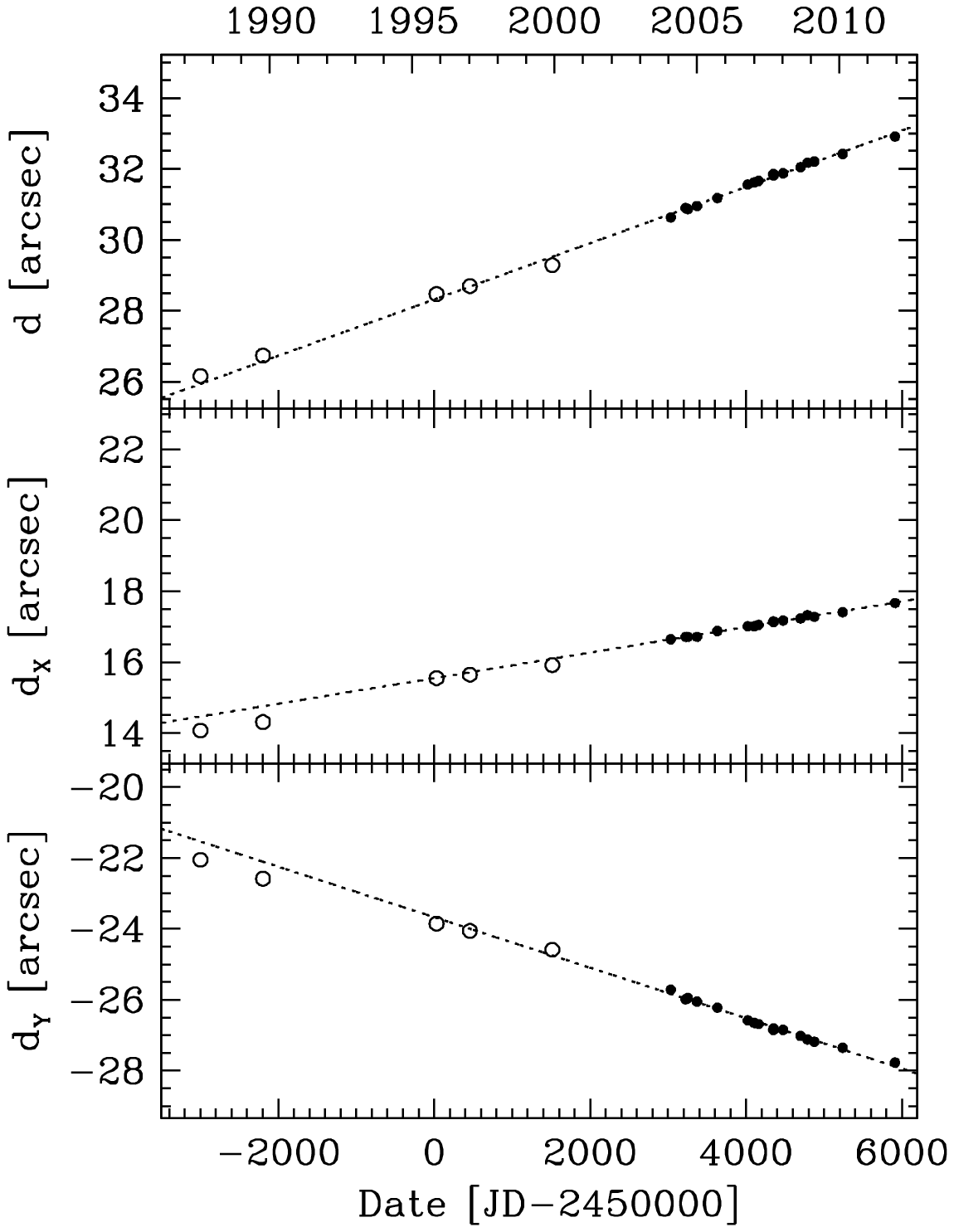}
\hspace{\fill}
\includegraphics[width=6.5cm]{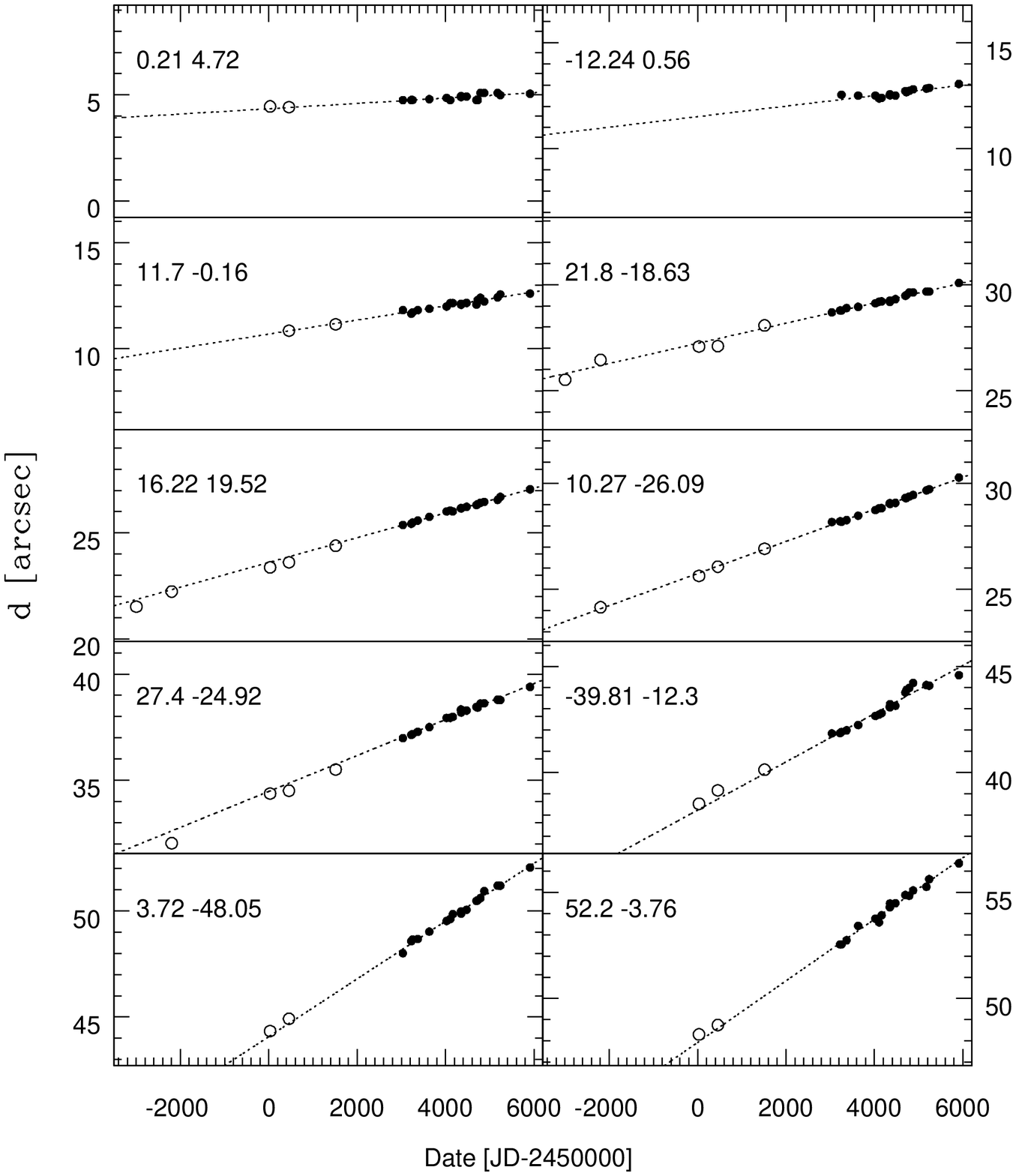}
\caption{Examples of proper motion calculations.  Left: The variation
  of distance $d$ with time and its components along the $X$ and $Y$
  axes ($d_{x}$ and $d_{y}$) for an individual knot. Right: additional
  examples, in which knots are ordered according to increasing proper
  motion (the range spanned by abscissae is the same for all graphs,
  $10"$). In all graphs filled circles are data from 2004 to 2011,
  open circles earlier archival data (not used for the $\mu$
  determination).  Error bars are smaller than or equal to the size of
  symbols).
\label{F-pm}}
\end{center}
\end{figure}

Our 19 carefully registered image obtained between 2004 and 2011
allowed a precise determination of the apparent expansion of the GK
Per remnant in the plane of the sky.  To calculate the proper motion
$\mu$ vector of individual knots the following approach was taken.  First,
the position of knots were measured by Gaussian fitting in the $X$ and
$Y$ directions.  This determines the apparent distance from the
central star ($d_{x}$, $d_{y}$, and $d$). Then $\mu$ was determined by
means of a least-squares fit of the positions at the different epochs,
as illustrated in the Fig.~\ref{F-pm}. This also provides the
direction of the the $\mu$ vector
($\alpha=\arctan(\mu_{x}/\mu_{y})$). In this way, 282 individual knots
were measured.  For all them, a straight line provides an excellent
fit to the variation of $d$ during the 7.9 yr lapse of time
considered.  Proper motion varies from $0".007$ yr$^{-1}$ to $0."53$
yr$^{-1}$.  When fixing the distance to the object, $\mu$ can be
transformed into velocity in the plane of the sky 
as follows: $v_{sky}\ [km\ s^{-1}] = 4.74 \cdot \mu\ [''\ yr^{-1}]
\cdot D\ [pc]$.

The WFC of the INT has well-defined and small geometrical
distortions
in the limited FOV covered by the remnant, which are robustly removed
during the astrometric registering of images. Therefore, errors in
the proper motion determination mainly depend on the goodness of the
fit of the position of the knots at the different epochs. The latter
depends on the knots shape and brightness and their behaviour, which
is rather stable. Consequently, as the proper motion error, we adopt
the formal error of the least squares fit: its average value is
$0".010\pm0".006$ yr$^{-1}$.

The archive images were not used for the proper motion calculations as
they constitute a less homogeneous set of observations, with generally
poorer resolution or astrometric properties, and different
filters. The archive images were used to search for signs of
acceleration/deceleration of knots in the past decade with respect to
the previous 20 yr. In this respect, we found no evidence for a
systematic acceleration/deceleration.

INT $H\alpha$+[N II] images were used for flux analysis. 
Similarly to measurements in radio \citep{tl_anup2005}, the total flux in optical has been linearly decreasing 
during the past decade with a rate of 2.6\% yr$^{-1}$.
However, the flux in different quadrants varies significantly. 
The NE quarter shows the shallowest decline and perhaps even a possible re-brightening 
in the past 3 yr. The SW quadrant is fading fastest. 

\subsection{Kinematical Ages}
\begin{figure}[t]
 \begin{center}
 \includegraphics[width=6cm]{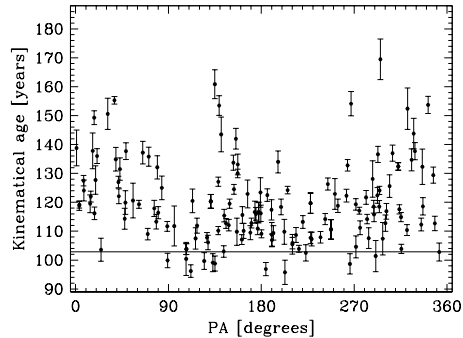}
\caption{Kinematical ages of 151 individual knots.
\label{F-age}}
 \end{center}
\end{figure}

Kinematical ages are calculated as follows in convenient units: $t~[yr] = d~["]~/~\mu~[''\ yr^{-1}]$. 
In order to focus on the directions close to the plane of the sky and minimize projection effects knots 
only with $d\ge35"$ and an estimated error on the age smaller than 8 yr were used. 
In this selection, 151 knots where left  (Fig.~\ref{F-age}). The ages vary from 96 to 170 yr. 
Mean value, weighted by errors, is $118\pm12$ yr. 
Horizontal line indicates the nova age 102.9 yr (from outburst until our first data) 
and shows that in general knots have suffered only a 
modest deceleration since their ejection. 

 \citet{tl_bode2004} predicted that the SW part of the remnant should be decelerated due to the 
interaction with the ancient bipolar nebula. The prediction is supported with the radio 
and X-ray observations. The SW part is a source of a non-thermal (synchrotron) radio emission 
\citep{tl_seaq1989} and has an highly asymmetric X-ray nebula \citep{tl_bal2005}. 
We can test this prediction when comparing the ages from different quadrants: 
NE $125\pm14$ yr, SE $115\pm11$ yr, SW $113\pm10$ yr, and NW $121\pm11$ yr. 
The lack of deceleration in SW part is obvious. There is no strong 
evidence for deviations from the circular symmetry. 
If anything, knots in the northern part seem to have suffered a stronger 
deceleration throughout the ejecta lifetime than those in the southern part. 
An attempt to explain the almost symmetric expansion can be made when considering 
hydrodynamical models \citep{tl_villa2012} and simplifying the GK Per to be a single star. Models show that once the bow-shock 
structure is formed, the subsequent stellar ejecta expands unperturbed inside the 
asymmetric outer cavity created by the interaction. The asymmetry observed in the radio 
and X-ray can be explained with the magnetized ISM interacting with the ancient bipolar nebula \citep{tl_sok1997}.

\section{Radial Velocities}
For radial velocity, $v_{rad}$, measurements the brightest optical emission line [NII] 6583~\AA\ was used. 
In total 217 knots were measured. 
Line profiles are generally broad, as also noted by \citet{tl_shara2012}, with an instrument-corrected FWHM 
of up to 200~km~s$^{-1}$. Knots toward the center of the nebula frequently have asymmetric shapes, 
with tails extending to velocities as large as 300~km~s$^{-1}$. Given the strong projection effects in central 
regions, it is not clear if these tails are intrinsic to the knots or caused 
by apparent superposition of more than one knot. $V_{rad}$ varies from $-989$ to $+967$~km~s$^{-1}$.
All measurements were corrected to the local standard of rest and with the systemic velocity from 
\citet{tl_bode2004} ($45\pm4$~km~s$^{-1}$). Typical errors are $8$~km~s$^{-1}$. 

\section{Kinematical Analysis}

\begin{figure*}[!b]
\begin{center}
\includegraphics[width=7cm]{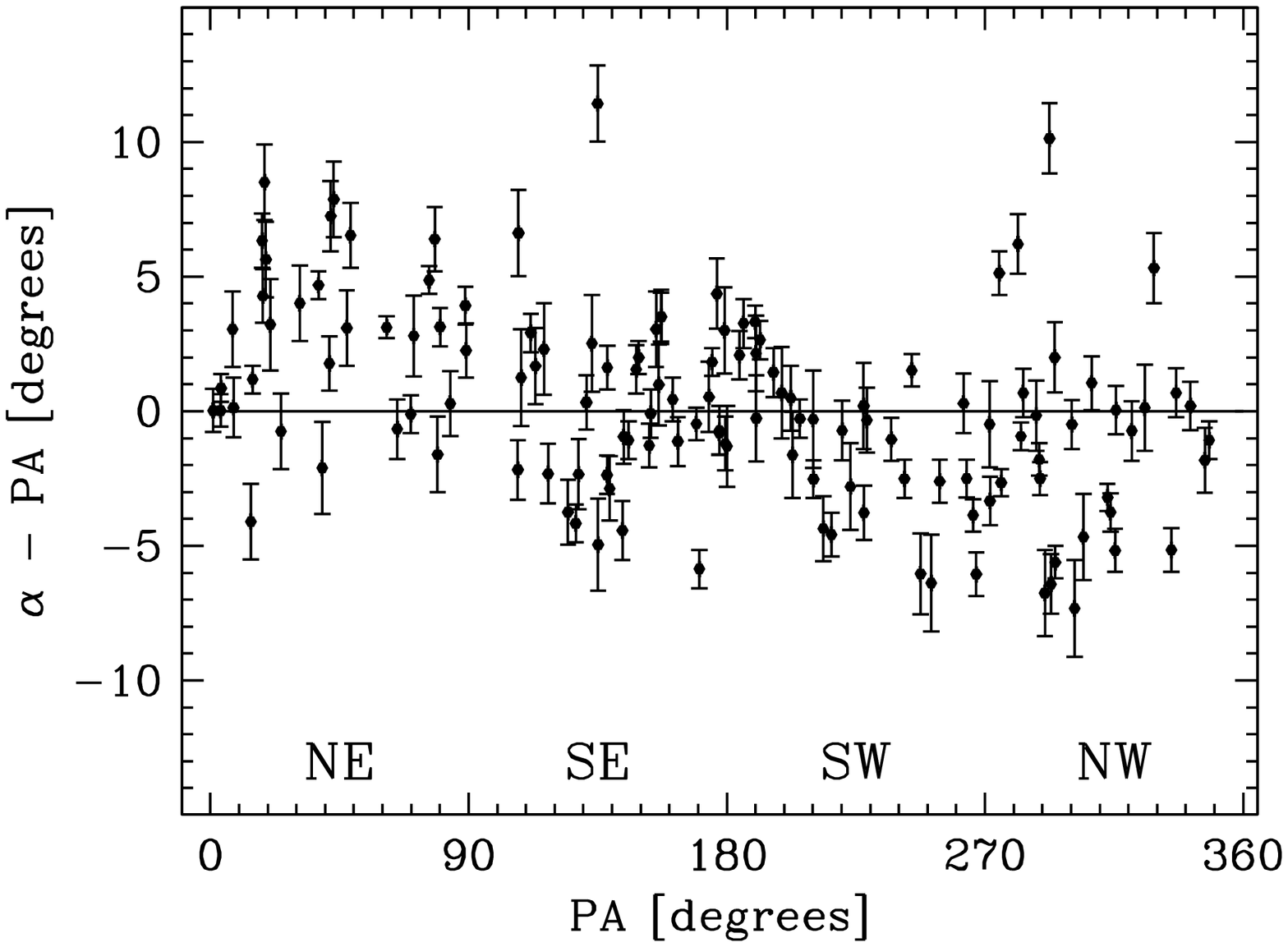}
\hspace{\fill}
\includegraphics[width=5.5cm]{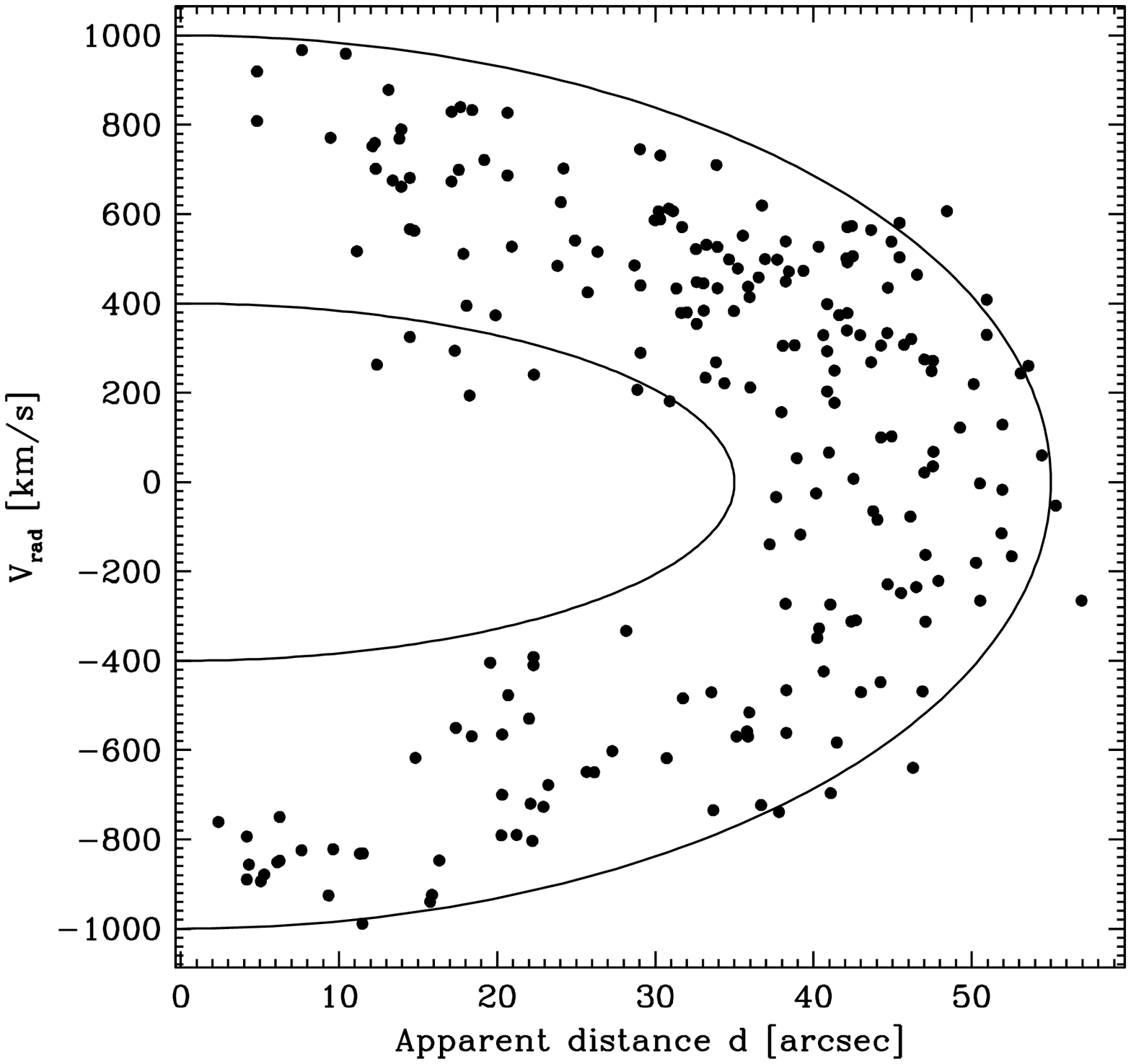}
\caption{Assumptions on quasi-spherical shell of GK Per. See text for more details.
\label{F-sph}}
 \end{center}
\end{figure*}

\begin{figure}[!ht]
 \begin{center}
  \includegraphics[width=6.5cm]{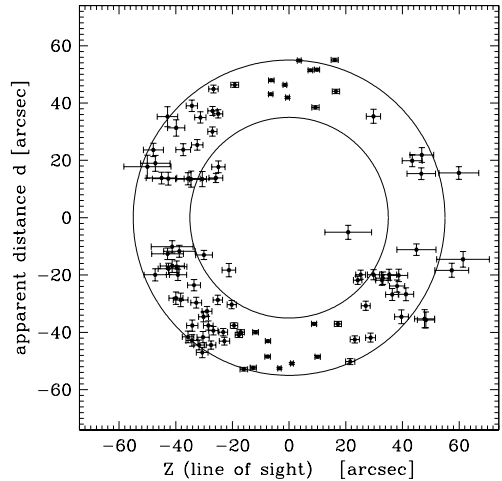}
  \caption{Representation of the depth of the remnant along the line of sight 
  \label{F-fold}}
 \end{center}
\end{figure}

In order to estimate the distance an assumptions about the real geometry of the GK Per was made.
Firstly, proper motion directions $\alpha$ were compared with position angles (PA) of knots in the plane 
of the sky (left panel Fig.~\ref{F-sph}) in order to know if the expansion is ballistic.
Velocity vectors are generally aligned along the radial direction, but a pattern of non-radial 
velocities is also observed. 
The plot of radial velocities  as a function of apparent distance from the central star 
together with the expected distance-velocity plot for spherical shells of radii of $35"$ and $55"$ 
and expansion velocities of $400$ and $1000$ km~s$^{-1}$(right panel Fig.~\ref{F-sph}) shows that most of the 
knots are confined within these limits, suggesting that the knots of the GK Per form a 
relatively thick shell expanding with a significant range of velocities. 
Considering the above mentioned and overall roundish morphology of GK Per we consider the remnant 
to be a quasi-spherical shell in which knots expand purely radially from the centre. 
With this in hand we can represent the depth of the nebula along the line of sight 
(Fig.~\ref{F-fold}). The observer is on the right side. 
99 knots with error on spacial distance $R < 10"$ are presented. 
The extension of the nebula along $Z$ depends on the adopted distance $D$. We assume that the distance 
value that better corresponds to an overall spherical shape of the nebula is our Òexpansion-parallaxÓ 
determination of the distance. The adopted value, estimated by visual inspection of the variation of the 
overall shape of the nebula as a function of $D$, is 400$\pm$30~pc. 

\section{Expansion Velocities}
Expansion velocity can be calculated as follows: $v_{exp} = \sqrt{v_{sky}^2+v_{rad}^2}$. 
Most knots have $v_{exp}$ between 600 and 1000~km~s$^{-1}$. Such a range of velocities 
could be due to the initial range of velocities or, due to the dynamical evolution of the two distinct 
ejecta components. The latter is also supported by the progressive circularization of the SW nebular 
edge and a relative brightening of the whole eastern side compared to the western part (see animation 
in \citet{tl_liim2012} Fig. 3). 

\section{Conclusions}
The ejecta of GK Per is a thick shell (nearly half of its outer radius) 
consisting of knots expanding with a
significant range of velocities. Since their ejection one century ago,
knots have suffered only moderate deceleration, with no clear signs of
dependence of the expansion rate on the position angle. 
The velocity vectors are generally aligned along the radial direction but a
symmetric pattern of non-radial velocities is also observed at
specific directions. The total $H\alpha$+[NII]  flux of the nebula seems
to be linearly decreasing with time, but with a significant difference
between the various quadrants. An improved kinematic
distance determination of $400 \pm 30$ pc was obtained.
These results raise some problems to the previous interpretations of
the dynamical properties of GK Per. In particular, the idea of
a strong interaction of the outflow with the surrounding medium in the
SW quadrant is not supported by our data, owing to the lack of
significant deceleration of the knots along this direction.

\acknowledgements TL would like to thank Tartu University Foundation, CWT Estonia AS, and conference organizers 
for their support in attending the conference.

\end{document}